\documentclass[12pt]{article}

\usepackage{times}
\usepackage{a4wide}
\usepackage{url}
\usepackage{epsfig}

\usepackage{mathpazo} 
\usepackage[scaled]{helvet} 
\usepackage{courier} 
\normalfont
\usepackage[T1]{fontenc}

\begin{document}

\title{Challenges and Directions for Engineering Multi-agent Systems\thanks{Dagstuhl Seminar 12342 (August 2012), ``Engineering Multi-Agent Systems'', organizers: J{\"u}rgen Dix, Koen V. Hindriks, Brian Logan, Wayne Wobcke.}}
\author{Michael Winikoff}
\date{29 August 2012}

\maketitle

\urlstyle{sf}


\begin{abstract}
In this talk I review where we stand regarding the engineering of multi-agent systems. There is both good news and bad news.  The good news is that over the past decade we've made considerable progress  on techniques for engineering multi-agent systems: we have good, usable methodologies, and mature tools. Furthermore, we've seen a wide range of demonstrated applications, and have even begun to quantify the advantages of agent technology. However, industry involvement in AAMAS appears to be declining (as measured  by industry sponsorship of the conference), and industry affiliated attendants at  AAMAS 2012 were few (1-2\%). Furthermore, looking at the applications of  agents being reported at recent AAMAS, usage of Agent Oriented Software Engineering (AOSE) and of Agent Oriented Programming Languages (AOPLs) is quite limited. This observation is corroborated by the results of a 2008 survey by Frank and Virginia Dignum. Based on these observations, I make five recommendations:
\begin{enumerate}
\item Re-engage with industry
\item Stop designing AOPLs and AOSE methodologies \ldots and instead \ldots
\item Move to the ``macro'' level: develop techniques for designing and implementing interaction, integrate micro (single cognitive agent) and macro (MAS) design and implementation
\item Develop techniques for the Assurance of MAS
\item Re-engage with the US
\end{enumerate}
\end{abstract}

\section{Introduction}

This paper is simply a written form of my presentation at the Dagstuhl seminar (12342, \textit{Engineering Multi-Agent Systems}, 20-24 August 2012). In it, I review where we stand, and, based on the review, propose a number of challenges and a number of research directions. 

This paper can be seen as an update to an earlier paper \cite{Winikoff2009}. As will be seen, some of the research directions identified in the earlier paper are still valid and important, and are proposed here as topics for future work. However, there are also some new directions and challenges identified here.

This paper is structured as follows. We firstly (briefly) review some good news (Section 2), including demonstrated applications, and progress made by the agent engineering community over the past decade or so. This is followed (Section 3) by the bad news: an assessment of our relevance, both in terms of the relevance of the overall field of Autonomous Agents and Multi-Agent Systems (AAMAS) to industry, and the relevance of the agent engineering sub-community to the rest of AAMAS. We then identify two research challenges (Section 4) before concluding with a final recommendation and a summary (Section 5).

\section{Good News}

There are three areas in which we have good news to celebrate. 

The first is demonstrated applications. There have now been many many demonstrated application of agents reported in the literature. These applications range from early work on fault diagnosis in the space shuttle, and aircraft landing sequencing at Sydney airport, to robocup soccer, electronic-commerce, and a wide range of more recent applications in the areas of information management, business process management, Holonic manufacturing, defence simulation, entertainment (e.g. black\&white, Massive), energy management and distribution, health, logistics, and emergency response management. 

Secondly, not only are there applications, but there has also been some work on quantifying the benefits of using agents (specifically BDI agents) for a range of applications. Statements such as the following are quite encouraging:

\begin{quote}``\textit{In a \textbf{wide range of complex business applications}, we show that the use of \textbf{BDI} [belief-desire-intention] technology incorporated within an enterprise-level architecture can \textbf{improve overall developer productivity by an average [of] 350\%}. For java coding alone, the increase in productivity was over 500\%''} \cite{Benfield2006} (emphasis added)
\end{quote}

However, we do need to bear in mind that the statement was made in a paper written by a commercial company, regarding their product. Unfortunately, I am not aware of any subsequent work that has considered the question of developer productivity using BDI agents. Consequently, we have to regard the above statements with a degree of skepticism. 

The third area for celebration is the progress that the sub-community concerned with engineering Multi-Agent Systems (MAS) has achieved over the past decade. Casting our minds back to the early 2000s, we might recall that agents were considered hard to develop. The challenges in teaching students, even advanced students, to design and implement agent systems included: the BDI model being difficult to understand and use; the lack of a mature and usable methodology (e.g. an AgentLink report noted that ``\textit{One of the most fundamental obstacles to large-scale take-up of agent technology is the lack of mature software development methodologies for agent-based systems.}''); the lack of textbooks, meaning that students had to rely on brief and somewhat inaccessible descriptions in research papers; and the lack of solid, well-developed (and supported!) tools. Thankfully, none of these are still an issue. We now have better ways of explaining the BDI model, there are a number of well-developed mature methodologies, there is a series of textbooks on the engineering of MAS, and there is a range of mature tools. Often we look forward, but sometimes one has to look back to see how far we have come.

\section{Bad News}

Unfortunately, there is bad news too. Specifically, I consider the following two questions:
\begin{itemize}
\item To what extent is industry involved in AAMAS as a whole?
\item To what extent are the methodologies, languages and tools developed by this sub-community used in the wider AAMAS community and by industry?
\end{itemize}

\subsection{Industry and AAMAS}

In order to assess industry involvement in AAMAS I use three metrics:
\begin{enumerate}
\item Industry sponsorship of the AAMAS conference;
\item Industry-based attendees of AAMAS; and
\item Industry-based authors of AAMAS papers.
\end{enumerate}

It should be noted that none of these are a perfect metric. What we really want to be able to assess is the extent to which industry sees agents (and AAMAS) as relevant and useful.  These metrics are all rough proxies, but, as we will see, even so, they are useful in getting some sense of industry involvement in AAMAS.

Let us begin with industry sponsorship at AAMAS. One can argue that to the extent that industry sees AAMAS as being relevant and of value, then they might be likely to consider sponsoring the conference. The table below shows that the number of industry sponsors has dropped from 7 in 2007, to 2 in 2011. In 2012 \textbf{\textit{none}} of the sponsors were industrial companies. Figure 1 shows the percentage of sponsors that are from industry. Although there is a clear downward trend since 2008, we can't conclude that this represents a loss of interest in agents. The beginning of this trend also coincides with certain other financial events that could explain why companies might be less likely to put money towards sponsoring a conference.

\begin{figure}
\begin{center}
\epsfig{file=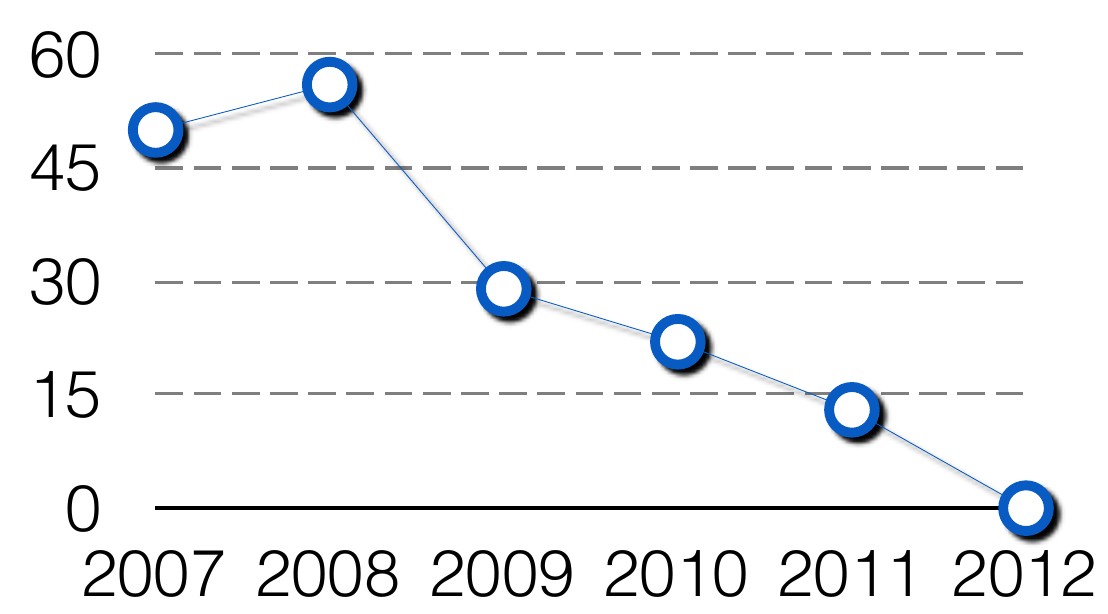,width=8cm,clip=}
\end{center}
\caption{Percentage of sponsors from industry}
\end{figure}

\begin{tabular}{|l|l|l|}
\hline
\textbf{Year} & \textbf{Total Sponsors} & \textbf{Industry Sponsors} \\ \hline 
2007 & 14 & 7 \\ \hline
2008 & 9 & 5 \\ \hline
2009 & 7 & 2 \\ \hline
2010 & 18 & 4 \\ \hline
2011 & 15 & 2 \\ \hline
2012 & 14 & 0 \\ \hline
\end{tabular}

We now turn to considering to what extent people from industry attend AAMAS. Unfortunately, I was only able to assess this for 2012. I took the list of registrants as of the second of May, comprising 655 registrants. I then searched for email addresses containing ``.com'' or ``.co'', on the assumption that someone from industry would likely have an address of this form (excluding gmail.com, yahoo.com, hotmail.com, and 163.com). However, as was subsequently pointed out to me, this assumption does not hold for all countries --- whilst some countries, such as the US, consistently use .com to denote industry, this is not true in all countries (for example .com is not used in France). This means that the following figures are in fact an under-estimate.

Of the 655 registrants, there were 14 from industry (2\%).  If we consider industry-based research labs (such as Microsoft Research, or IBM Research) as being, in some sense, not really industrial, then the number drops further.

Finally, we consider to what extent industry-based people are authors on AAMAS papers. Specifically, I looked at the industry track (2010) and the innovative applications track (2012), since that it arguably where industry work is most likely to be found.

Looking at the 2012 innovative applications track, there were 7 full papers. Of these, only one had any authors who were not from academia. The remaining 6 papers had authors who were all academics. The 7th paper was a collaboration between academia and the US coast guard, and included authors from the coast guard. 

In 2012 there were also 7 short papers in the innovative applications track. Of these, five solely had academics as authors, and the remaining two had both academics and industry people as authors.

Considering now 2010, before the industry track was changed to be the innovative applications track, there were 10 full papers. Eight of these had industry authors (either exclusively, or in combination with academics). 

To summarize:
\begin{itemize}
\item Industry sponsorship of AAMAS has been declining since 2008, although, given the financial context, we cannot conclude solely from this that industry interest is declining.
\item The number of industry-based attendees at AAMAS 2012 is estimated to be low. 
\item The vast majority of innovative applications papers (2012) did not have industry authors, but the industry track (in 2010) did have a significant degree of industry authorship.
\end{itemize}

This leads directly to the first of five recommendations: 
\begin{center}
\textit{\textbf{
1. Re-engage with industry}} \end{center}

We might want to reconsider the change from industry track to innovative applications track. Another point is that it would be desirable to encourage industry to attempt to quantify the benefits (and costs!) of agent technology, i.e. replicating and building on the work of Benfield \textit{et al.} \cite{Benfield2006}

Additionally, we need to engage not just with industry, but also with other relevant communities. For example, the service-oriented computing community, which faces similar issues, and has a tendency to re-invent wheels that we have already developed.

\subsection{Relevance of the Engineering MAS sub-community}

We now focus in on the sub-community concerned with Engineering Multi-Agent Systems, and ask to what extent is our work relevant? Specifically, to what extent are the methodologies and agent platforms that we have developed being used?

Fortunately, answering this question (in a preliminary way) is easy. Back in 2008, Virginia and Frank Dignum did a survey\footnote{J{\"o}rg M{\"u}ller is currently running a similar survey, but results are not yet available.} which collected precisely this information, with the results being included in a 2010 paper \cite{Dignum2010}. They received 25 responses to their survey, of which 84\% were from academia, 8\% from industry, and the remaining 8\% from other sources. Of the responses, 64\% were from the EU, and 28\% from North America. 

Figure 2 is based on data from their paper. It shows that 40\% of the responses (i.e. 10 of the 25 responses) did not use any agent-oriented platform. Of the remaining 60\%, five (20\%) used Jade/Jadex, and one project used JACK (4\%). The remaining 9 projects used a range of agent platforms, including two proprietary ones, and one using Brahms. The remaining platforms are not really what this community would consider (cognitive) agent platforms. Furthermore, it's not clear to what extent ``Jade/Jadex'' is really just JADE (i.e. not a cognitive agent platform). To summarize, only 28\% of the projects used an agent platform that was developed by our community (i.e. Jade(x), JACK, Brahms), and it may be that as few as 8\% (JACK and Brahms) used a \textit{cognitive} agent platform. It is also noteworthy that a number of well-known agent platforms are not represented here --- Jason and 3APL are notable by their absence! (the GOAL implementation didn't exist in 2008)

\begin{figure}[h!]
\begin{center}
\epsfig{file=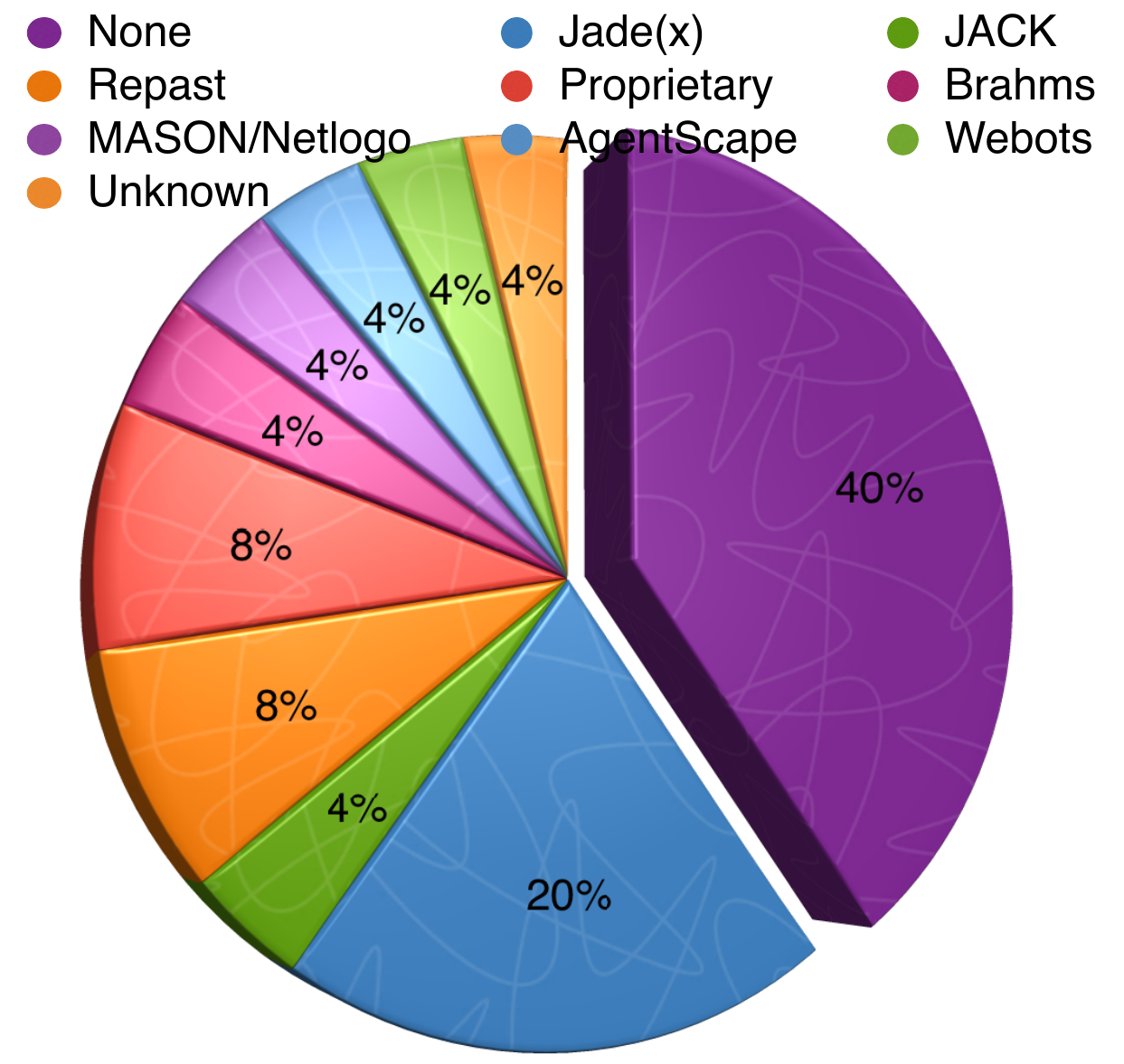,width=8cm,clip=}
\end{center}
\caption{AOPL usage (from Dignum \& Dignum 2008 survey) }
\end{figure}

We now consider methodologies.  The survey results (see Figure 3) indicates that 56\% of the projects did not use an AOSE methodology. OperA, Prometheus, and INGENIAS were each used by one project, and Gaia and O-MaSE were each used by two projects. A further 4 projects used an AOSE methodology, but didn't indicate what it was.

\begin{figure}[h!]
\begin{center}
\epsfig{file=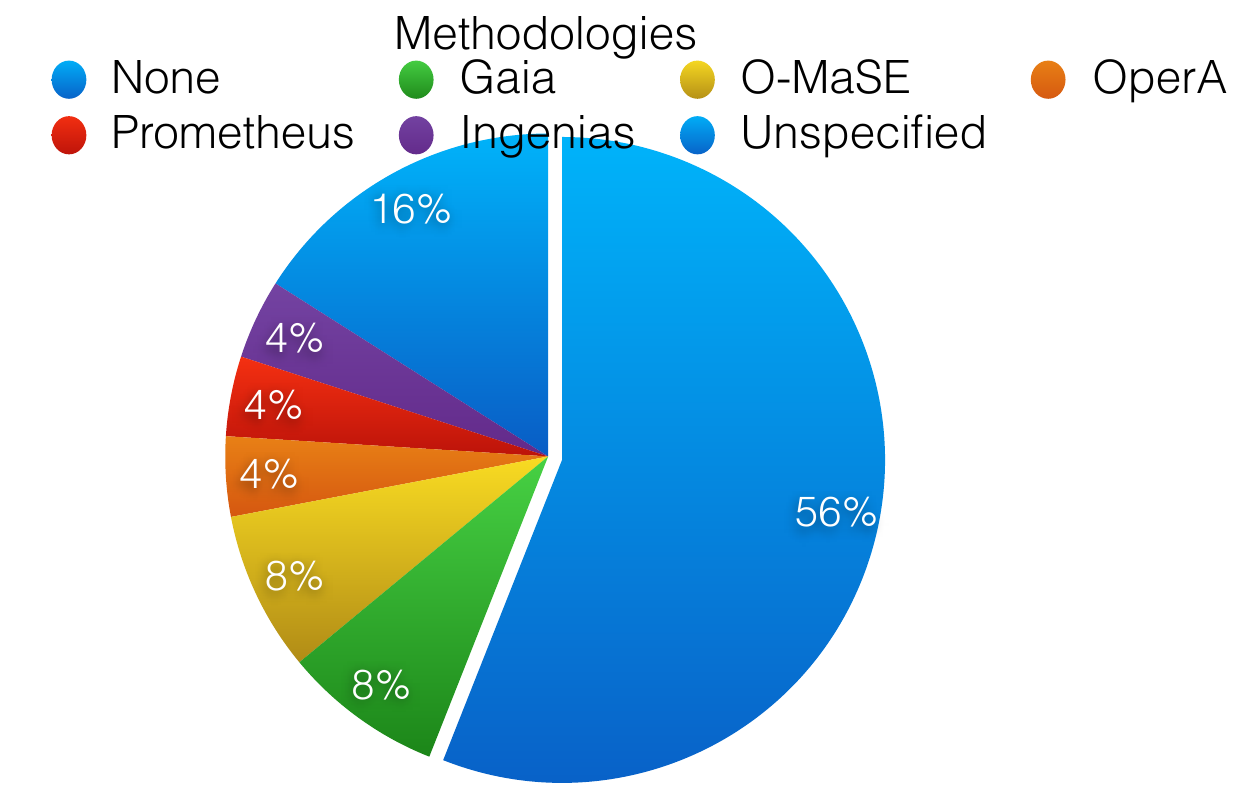,width=10cm,clip=}
\end{center}
\caption{AOSE methodology usage (from Dignum \& Dignum 2008 survey) }
\end{figure}
It does need to be noted that the sample size (25) is fairly small, and that it may not be representative (it certainly isn't comprehensive!). To compensate for this (at least somewhat) we also considered the AAMAS 2012 and 2010 applications papers. To what extent did these use AOPLs or methodologies? 

In 2012 none of the applications described in any of the 7 full papers used an agent oriented programming language (AOPL), or an agent-oriented software engineering methodology (AOSE). Of the 7 extended abstracts, one mentioned cognitive agent architecture, and one mentioned BDI.  In 2010, there was no evidence of AOPL usage, and limited AOSE usage.

It therefore appears that AOPL and AOSE usage is quite limited. Looking at the 2012 and 2010 applications papers it is clear that the applications in question are concerned with coordination aspects, and do not require goal-based agents. For example, one paper described an application that used game theory to schedule coast guard patrols. Another paper considered energy conservation using MDPs. Another paper looked at the smart grid. For these sorts of applications the challenges are around coordination, not planning.

This leads us to our second (and most controversial!) recommendation:
\begin{center}
\textit{\textbf{
2. Stop designing (new?) programming languages and methodologies.}} \end{center}
But if we are to no longer design AOPLs and AOSE methodologies, what should we do instead?

\section{Challenges and Directions}

I will propose two directions for work that I believe are significant issues which hinder the adoption of agent technology: agent interaction design and implementation, and assurance.

\subsection{Engineering Agent Interactions}

Much of the work that has been done on agent-oriented programming languages has focussed on the agent architecture, and on features that relate to a \emph{single} agent. For example, the BDI execution cycle is a single-agent reasoning cycle. Although all AOPLs support the development of multi-agent systems, not just of single agents, they almost invariably do so simply by providing means for agents to send and receive messages. 

Turning to AOSE methodologies, they tend to define agent interactions in terms of message-centric interaction protocols, often expressed as AgentUML (AUML). The problem is that these protocols tend to over-constrain the interaction patterns, and do not leave the agents room to be autonomous or to exploit their flexibility and robustness when interacting with other agents. 

What we need to do is to develop better techniques for designing and implementing agent interactions. These techniques should exploit the ability of individual agents to be flexible and robust, allowing interactions to also be more flexible and robust. 

There has been some work in this area already. For example, the use of social commitments to design interactions; work on teamwork that frames agent interaction in terms of the execution of ``joint goals'', or of ``team plans''; and work on ``interaction goals''. There has also been work on extending AOPLs with representations for organisations, and associated concepts (such as roles). However, this whole area of work is still immature and is not ready for real use on real applications.

Finally, we need to move away from the direct use of messages to implement agent interactions. A message send can be seen as a transfer of control: the thread of execution continues in the recipient. Indeed, we can see explicit message sending and receiving as being the agent equivalents of ``gotos'' and ``labels''. Worse, if the message is asynchronous then we also introduce concurrency, along with all the challenges that accompany it. There is no doubt that messages between agents are needed at the implementation level, but I argue that we need to develop higher level abstractions that hide the explicit use of messages, in exactly the same way that using transactional memory hides the explicit use of locks to avoid race conditions, making certain errors impossible to make, and raising the level of programming abstraction. 

\subsection{Assurance}

\begin{quote}
``\ldots \textit{systems exhibiting emergent behavior [are] ...the equivalent of a specification that says `surprise me'.}''  --- Lou Mazzucchelli, in \cite{Odell2007}
\end{quote}

One of the strengths of agents and multi-agent systems is that they are able to deal with a range of situations, balancing proactivity and reactivity as needed. However, as the above quote illustrates, this raises an essential question: how can I be confident that my system will work ``reasonably'' in all situations?

Techniques for obtaining assurance that a system will not misbehave are clearly important, as the following quote illustrates:
\begin{quote}
``\textit{The verification of such complex software is \textbf{critical to its acceptance by science mission managers.}}'' \cite{Havelund2000} (emphasis added)
\end{quote}

Although this need for assurance is not new, what is new is the ability of agent systems to exhibit complex, emergent, and sometimes surprising behaviour:
\begin{quote}
`` \ldots \textit{validation through extensive tests was mandatory \ldots However, the task proved challenging for several reasons. First, agent-based systems explore realms of \textbf{behaviour outside people's expectations and often yield surprises \ldots}}'' \cite[section 3.7.2]{Munroe2006} (emphasis added)
\end{quote}

The challenge is to be able to balance the need for adaptability with the need for predictability, and this is still an outstanding issue:
\begin{quote}
``\textit{Establishing appropriate trade-offs between adaptability and predictability. Creating systems able to adapt themselves to changing environments, and to cope with autonomous components, may well lead to systems exhibiting properties that were not predicted or desired. \textbf{Striking a balance, appropriate to the specific application domain, between adaptability and predictability is a major challenge, as yet unresolved} either theoretically or practically. Associated with predictability is the requirement for practical methods and tools for verification of system properties, particularly in multi-agent systems that are likely to exhibit emergent behaviour}''  \cite[Page 78]{Luck2005} (emphasis added)
\end{quote}

We note that this issue is broader than just verification against a specification \cite{Winikoff2010}. For instance, specifications may be incomplete, and some aspects of desired behaviour may be difficult to specify. For example, in temporal logic it is easy and natural to specify that a certain condition is ``eventually'' achieved. However, ``eventually'' may be too weak, and a deadline of ``reasonably soon'' may be more appropriate. Furthermore, requiring that a property ``eventually'' hold can also be too strong in that it fails to recognise that in some situations the desired property may be unachievable. For instance, in allocating the delivery of packages in a logistics application, a key bridge may be unavailable. 

A detailed discussion of various techniques to be considered for obtaining assurance is beyond the scope of this document. We therefore merely note that areas for work include correctness proofs, model checking, various forms of testing, and the use of safety cases.

\section{Summary}

Before proceeding to summarize the recommendations, I would like to make one more observation, which leads to a final recommendation. 

The community concerned with engineering MAS is predominantly a European community. Looking at the attendees of this Dagstuhl workshop the vast majority are from Europe. We have a single Canadian, a single American (of Dutch origin!), and a handful of people from outside Europe (some of whom can be seen as European in terms of their training, style of work, and connections). 

This state of affairs is undesirable: we need to re-engage with our North American colleagues. Simply put, a significant part of the AAMAS community is from North America (e.g. in AAMAS 2012 there were 92 papers from North America, 122 from the EU, 32 from Asia, and 36 from the rest of the world [Australia, Brazil, Chile, Israel, UAE]). By not engaging with the North American part of the AAMAS community we are missing out on working with a large part of the community, and are also missing out on opportunities to raise awareness of our work, so that it can be used where it is relevant. Finally, looking more generally at computing, it is noteworthy that (according to a recent ranking\footnote{\url{http://www.shanghairanking.com/SubjectCS2012.html}}) of the top 33 computer science departments 25 are in North America (with the remainder being in Israel (4), China (2), and Europe (2)).

To summarise, our five recommendations are:
\begin{enumerate}
\item Re-engage with industry
\item Stop designing AOPLs and AOSE methodologies \ldots and instead \ldots
\item Move to the ``macro'' level: develop techniques for designing and implementing interaction, integrate micro (single cognitive agent) and macro (MAS) design and implementation
\item Develop techniques for the Assurance of MAS
\item Re-engage with the US
\end{enumerate}

\end{document}